\documentclass[aps,prl,twocolumn,groupedaddress,showpacs]{revtex4}
\usepackage{graphicx}
\usepackage{dcolumn}
\usepackage{epsfig}
\usepackage{amssymb}
\usepackage{amsmath}
\usepackage{rotating}
\usepackage{color}

\def\ie{{\it i.e.} }

\begin{document} 

\title{Superhydrophobic coatings against the teapot effect: capillarity effects and dripping}
\title{Beating the teapot effect with superhydrophobic coatings}
\title{Lotus versus teapot effect: 1:0}
\title{Beating the teapot effect}

\author{Cyril Duez$^\dag$, Christophe Ybert$^\dag$, Christophe Clanet$^\ddag$, Lyd\'eric Bocquet$^\dag$}
\email{lyderic.bocquet@univ-lyon1.fr}
\address{$^\dag$ Laboratoire PMCN, Universit\'e Lyon 1, UMR CNRS 5586, 69622 Villeurbanne, France \\
$^\ddag$ LadHyX, Ecole Polytechnique, UMR CNRS 7646, 91128 Palaiseau, France}




\begin{abstract} 
We investigate the dripping of liquids around solid surfaces in the regime of inertial flows,
a situation commonly encountered with the so-called ``teapot effect''.
We demonstrate that surface wettability is an unexpected key factor in controlling flow separation and dripping, the latter being completely suppressed in the limit of  superhydrophobic substrates. This unforeseen coupling is rationalized in terms of a novel hydro-capillary adhesion framework, which couples inertial flows to surface wettability effects. 
This description of flow separation successfully captures the observed dependence on the various experimental parameters --wettability, flow velocity, solid surface edge curvature--.
As a further illustration of this coupling, a real-time control of dripping is demonstrated using electro-wetting for contact angle actuation. 

\end{abstract}

\maketitle

Over the recent years, the development of  super-hydrophobic materials, exhibiting
the so-called Lotus effect, has stirred up the physics of surfaces 
\cite{quere,Blossey}. 
Their exceptional water repellency results from the combination of bare
hydrophobicity and micro- or nano- structures decorating the solid surface.
These materials
have triggered research, leading to the discovery 
of unforseen phenomena,  like bouncing drops \cite{Richard} or big splashes of impacting bodies\cite{Duez}, the exploration of which is still in its infancy.
Generally, their extreme wetting behavior 
raises the question of its potential impact on the {\it dynamics} of fluids at large scales.
The question is however far from obvious, as surface effects are only expected to affect fluid dynamics at small scales, while
inertia rules the world of large scales hydrodynamics. 
This is usually quantified by dimensionless numbers, Reynolds ${\rm Re}=\rho U a/\eta$, but also Weber number, ${\rm We}=\rho U^2 a/\gamma $ (with $\gamma$ a typical surface energy, $\eta$ the shear viscoity, $a$ a typical length scale, $U$ a velocity, $\rho$ the mass density). In the large scale flow regime, ${\rm We}\gg 1$, surface energies are negligeable and surface properties are not expected to be a relevant factor.


\begin{figure}[t]
\begin{center}
\includegraphics[width=8cm,height=!]{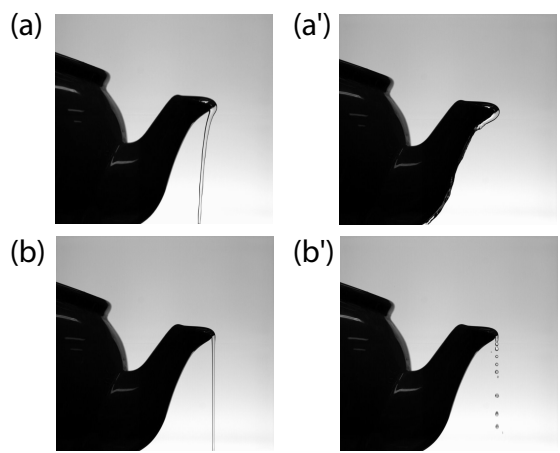}
\caption{{Beating the teapot effect with a superhydrophobic coating.} {\it Top}: water flow under the spout of an (hydrophilic) teapot, exhibiting a bending of the streamlines  ($a$), and dripping as the water flow decreases ($a^\prime$).
{\it Bottom}: In contrast, a teapot with a spout coated by a superhydrophobic coating (here black soot) fully avoids dripping ($b$ and $b^\prime$).
\label{fig1}}
\end{center} 
\end{figure}

Nevertheless,  we 
unveil in this work unexpected coupling channels 
 that 
connect these different worlds. 
We 
explore here this question on the dripping phenomenon, which is commonly known as the ``teapot effect''. This is examplified in Fig. \ref{fig1}: a ``rapid'' water flow poured from a teapot is shown to bent and finally drip along the teapot as the flow decreases, Fig. \ref{fig1}-(a-a'). Now, as shown in Fig \ref{fig1}-(b-b'), treating the spout of the teapot with a superhydrophobic coating (here, black soot) fully eliminates this effect: superhydrophobic coatings indeed beat the teapot effect~! More fundamentally, this result points to an {\it a priori} unexpected link between water repellency and large scale flows.

However, contradictory interpretations have been given to the teapot effect, especially concerning the role of surface adhesion:  
for rapid flows -- ${\rm We}, {\rm Re} \gg 1$ --
surface adhesion is not expected to play a role as mentionned above, and dripping is then interpreted in terms of
bending of streamlines and flow separation \cite{Walker,Reiner,Keller,VDB,POF09}. 
On the other hand, the work by Kistler and Scriven \cite{Scriven}
pointed an influence of wettability on the dripping phenomenon, however for a highly viscous fluid, 
\ie
associated with a rather small Reynolds number, a regime where capillary effects are indeed expected.
These frameworks however cannot account for the observation of Fig. \ref{fig1}, which points to a direct effect of wettability even in the ``fast flow'' regime.

\begin{figure}[t]
\begin{center}
\includegraphics[width=8cm,height=!]{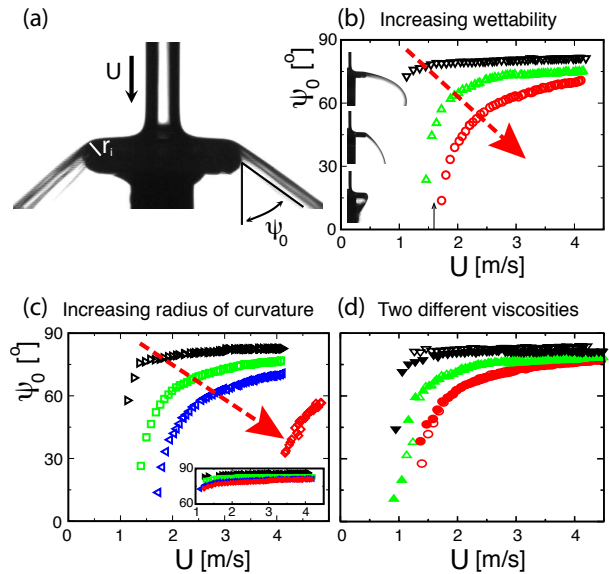}
\caption{
Flow ejection versus wettability, geometry and viscosity. 
{\it (a)} Experimental set-up: a liquid jet with velocity $U$ impacts a solid  surface, characterized by its wettability $\theta_0$ and the radius of curvature $r_i$ of its edge. The fluid ejects at an angle $\psi_0$, measured from the detailed image analysis of the fluid surface.  {\it (b)} Velocity dependence of the ejection angle $\psi_0(U)$ for various impacters with increasing wettability, for a fixed radius of curvature ({$r_i=1$mm}): from top to bottom, $\theta_0=175^\circ$, $115^\circ$, $10^\circ$. 
 In the inset, images of the deflections are shown for various wettabilities at a given velocity $U={1.6}5\pm 0.05$ m.s$^{-1}$ (indicated as an arrow on the bottom axis of the main panel). 
 {\it (c)} Velocity dependence of the ejection angle $\psi_0(U)$ for various impacters with increasing radius of curvature for a given wettability $\theta_0=10^\circ$: from top to bottom {$r_i \simeq 0.03$mm}, $0.5$mm, $1 mm$, 2 $mm$. As an inset, similar plots are shown for the case of a superhydrophobic coating $\theta_0=175^\circ$ (same symbols), showing no dependence on the radius of curvature for this extremely water repellent surface.  {\it (d)} Velocity dependence of the ejection angle $\psi_0(U)$ obtained for liquids with two different viscosities: water $\eta_w=1$mPa.s (filled symbols); and a water-glycerol mixture, with a doubled viscosity $\eta_{w/g}=2$mPa.s (open symbols).
From top to bottom,  results are shown for various impacters with  $\theta_0=175^\circ$, $115^\circ$, $10^\circ$ and {$r_i=0.5mm$}. The results for the ejection angle $\psi_0(U)$ are found to superimpose for the two liquids. 
\label{fig2}}
\end{center} 
\end{figure}

To get further insight into the physical mechanisms at the origin of this phenomenon, we have performed a systematic study of the ejection and dripping of liquids in a controlled geometry, with varying surface properties
and geometrical characteristics. 
The setup is sketched in Fig. \ref{fig2}: a water jet with velocity $U$ --  typically from 1 to 5 m.s$^{-1}$ in our study -- and diameter $D$  -- here $D=4$mm -- impacts and spreads over a solid surface (the 'impacter') with a given wettability and radius of curvature of its edge (``the spout'').

Impacters consists in horizontal disks of diameter $D_i=15$mm, ended by a curved edge characterized by its radius of curvature $r_i$. We studied four different geometries with $r_i$ ranging between $2$mm and $0.03$mm ($r_i=2$mm, $1$mm, $0.5$mm, $0.03$mm).
The wettability of the impacters is tuned by using different chemical processes,
leading to a static (advancing) contact angle $\theta_0$ ranging between 10$^\circ$ up to 175$^\circ$ 
(superhydrophobic coating).  A contact angle of $\theta_0=78\pm 5^\circ$ is obtained for cleaned, native purum Aluminium (Al 1050) impacters. A treatment in a UV-O$_3$ reactor lead to a strongly hydrophilic impacter with a contact angle decreasing to $\theta_0=10\pm5^\circ$. Hydrophobic impacters were obtained by
grafting fluorosilane chains (perfluoro-octyltriethoxysilane) on the aluminium surface, leading to
$\theta_0=115\pm5^\circ$. Finally,
superhydrophobic impacters were obtained using galavanic deposition on purum copper (Cu-OF) impacters \cite{Larmour}.

For these different impacters, we measured the ejection angle $\psi_0$ ( Fig. \ref{fig2}) versus
fluid velocity $U$, for various geometries ($r_i$) and wettabilities ($\theta_0$).
Our results are gathered in Fig. \ref{fig2}. 
Altogether, 
these experiments fully confirm the previous observation in Fig. \ref{fig1}: the wettability of the surface has a key influence on the ejection of the fluid film from the surface, as highlighted here by the strong impact of the wettability of the  impacter on the ejection angle $\psi_0$, Fig. \ref{fig2}-b.
Superhydrophobic impacters 
strongly eject the liquid film, thereby avoiding dripping.
Furthermore, as one intuitively expects,  the radius of curvature of the impacter is found to have a strong influence on ejection,  Fig. \ref{fig2}-c. But again, superhydrophobicity is found to prevail over this geometrical parameter, as demonstrated in the inset of Fig. \ref{fig2}-c.

Viscosity is found {\it not} to be a relevant parameter for the present ``fast flow'' experiment, as one expects in the inertial regime: as shown in  Fig. \ref{fig2}.d, the ejection does not depend on the viscosity of the fluid.
 This observation implicitly dismisses a visco-capillary origin of the phenomenon. 
This is consistent with the rather large Reynolds number characterizing the flow, $Re_\ell=U\,\ell/\nu$ (
$Re_D\sim 10^4$ for $\ell=D$, the initial jet diameter, while $Re_e\sim 500-10^3$ with $\ell=e_0$, the film thickness $e_0$). Note also that gravity effects, as quantified by a Froude number $Fr=g\ell/U^2$ with $g$ the gravity constant, 
play a negligeable role here.

\begin{figure}[t]
\begin{center}
\includegraphics[width=8cm,height=!]{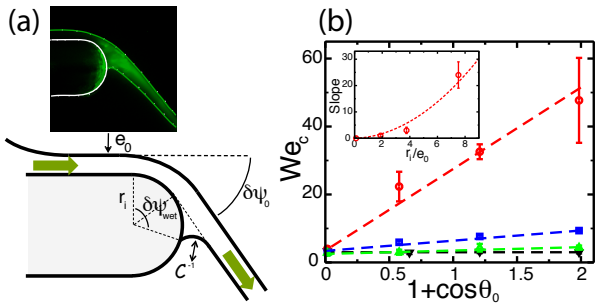}
\caption{ 
{\it (a)} Details of the flow around the edge of the impacter. The fluid film with thickness $e_0$ bends around the edge of the impacter, with radius of curvature $r_i$; $\delta\psi_0={\pi \over 2}-\psi_0$ is the deflection angle and $\delta \psi_{\rm wet}$ the angular range of the curved wetted area ($\delta\psi_{\rm wet}=\delta\psi_0+\delta\psi_{\rm menisc.}$ with $\delta\psi_{\rm menisc.}$ the angular width of the meniscus).  {\it Inset (top)}: experimental picture of a cut of
the liquid interface under flow, obtained using a laser sheet and a fluorescent dye. Solid lines are a guide for the eye of the various interfaces. {\it (b)} Threshold Weber number for dripping, ${\rm We}_c=\rho U_c^2 e_0/\gamma$, versus wettability $1+\cos\theta_0$, for various radius of curvature of the impacter's egde: from bottom to top
$r_i\simeq$ {$0.03$mm}, $0.5$mm, $1$mm, $2$mm. Dashed lines are linear fits. {\it Inset:}  plot of the slope of the fitted linear regression versus radius of curvature $r_i$. The dashed line is a fit according to the expected scaling 
for the slope as ${\cal R}\, r_i/e_0^2$, with ${\cal R}=r_i+e_0/2$, in Eq.(\ref{Wec}). 
The film thickness
$e_0$ is estimated using Bernoulli equation as $e_0\simeq D^2/4D_{\rm imp}$ with $D$ the liquid jet diameter and $D_{\rm imp}$ that of the impacter \cite{Clanet}.
 \label{fig3}}
\end{center} 
\end{figure}

Finally, a threshold for dripping can be identified experimentally: below a minimum velocity $U_c$, the
liquid is not ejected from the impacter anymore but drips along the spout, see insets in Fig.\ref{fig2}-b. 
We gather in Fig. \ref{fig3}-b the results for the threshold velocity  -- here plotted in terms of a dimensionless Weber number -- as a function of the wettability of the impacter and its radius of curvature. 
As intuitively expected, dripping occurs more easily for spouts with thicker edges. On the other hand, superhydrophobic coatings avoid dripping whatever the radius of curvature of the edge, in line with our previous observation  in Fig.\ref{fig1}. Furthemore
these plots suggest a linear dependence of the threshold Weber number, ${\rm We}_c$, versus 
$1+\cos\theta_0$, with a prefactor depending strongly on the impacter's radius of curvature $r_i$
(inset of Fig. \ref{fig3}-b).


Altogether these experimental observations points to the two key parameters controlling flow separation: the curvature of the ``spout'' and more unexpectedly its wettability. Dripping is fully avoided in the limit of sharp edges or superhydrophobic surfaces. However the underlying 
physical mechanism remains to be discovered: how to couple fast (inertia dominated) flows with wettability effects ? 


As we now show, these experimental observations can be rationalized in terms of a novel
``hydro-capilllary'' adhesion mechanism. The key point underlying the proposed mechanism is the existence
of a capillary meniscus connecting the flow to the spout's surface. As sketched in Fig. \ref{fig3}-a,
the liquid interface should connect the solid surface with an imposed angle given by the wetting contact angle $\theta_0$, thereby constraining its global geometry. As an illustration, we show in Fig. \ref{fig3}-a (top) an experimental picture of a cut of the liquid interface under flow, obtained using a laser sheet and a fluorescent dye.

It is this unforeseen detail which fully controls the fluid ejection and flow separation, thereby allowing to connect the inertial flow to surface capillary details. 
The overall picture runs accordingly as follows: the liquid sheet ejection results
from the balance between (i) the centrifugal force resulting from the inertia of the fluid flow, and (ii)
a hydro-capillary adhesion force.
The latter takes its origin in the pressure drop associated with the 
bending of the streamlines; this pressure drop acts over a ``wetted'' area, the size of which is fixed by the geometrical constraints set by capillarity: static contact angle $\theta_0$ and spout radius of curvature $r_i$. 
This adhesion
contribution will be accordingly calculated following a direct analogy to classical adhesion theory \cite{Israel},
here applied in the context of fast flow dynamics. 

Let us formalize this picture. 
We first consider the regime of  
large fluid velocity $U$, 
where the deviation $\delta\psi_0={\pi\over 2}-\psi_0$ of the water sheet is small. Conservation of momentum shows that the centrifugal force to bend
the streamlines is, when projected on the horizontal direction $x$: $F^x_{\rm cent}=\rho_w U^2 e_0(1-\sin\psi_0)\approx{1\over 2}  \rho_w U^2 e_0\, \delta\psi_0^2$ (per unit axisymetric length).
Now, 
to maintain a fixed deviation, 
this centrifugal force should be compensated by an attractive force to the surface.
Here we argue that this adhesion force takes its origin in the (negative) pressure drop $\Delta P$ induced by the bending of the streamlines \cite{Guyon}: $F_{\rm adh} \sim \Delta P \times {\cal A}_{\rm wet}$, with
${\cal A}_{\rm wet}$ the curved wetted region, as depicted in Fig.~\ref{fig3}
(${\cal A}_{\rm wet}\approx r_i \delta\psi_{\rm wet}$, per unit axisymetric length).
Projected along the horizontal, this yields $F_{\rm adh}^x \approx {\cal A}_{\rm wet} \times \Delta P \times{\delta \psi_{\rm wet} \over 2} $. 

Now, a key point of the hydro-capillary picture is to connect the wetted area ${\cal A}_{\rm wet}$, and thus the adhesion force,  to the 
location and geometry of the capillary meniscus. 
This is a classical problem in capillarity \cite{Israel}, with {\it e.g.} applications in adhesive granular materials \cite{Bocquet98}, and we follow this standard line of description here
\cite{Israel}.
Accordingly, the lateral size of the meniscus characterized by $\delta\psi_{\rm menisc}$ is set geometrically by the contact angle $\theta_0$ 
and the pressure drop $\Delta P$ fixing the curvature ${\cal C}=\vert \Delta P\vert /\gamma$ of the meniscus: 
$\delta\psi_{\rm menisc}^2= 2{\cal C}^{-1}/r_i\times (1+\cos\theta_0)$ \cite{Israel}. 
This leads to the final expression for 
the curved wetted area ${\cal A}_{\rm wet}$ on which the pressure drop applies 
as $\delta\psi_{\rm wet}=\delta \psi_0+\delta\psi_{\rm menisc}$.

The last part is to evaluate the pressure drop $\Delta P$, which originates in the bending of the streamlines. 
Denoting ${\cal R}$ the radius of curvature of the flow streamlines, then one expects $\Delta P \approx - \rho_w U^2 e_0/{\cal R}$ \cite{Guyon}. Typically ${\cal R}$ may be estimated as an averaged radius over the fluid film thickness $e_0$, which we write ${\cal R}= r_i+ \alpha e_0$, with $\alpha \approx {1\over 2}$.
%

Gathering these different results, the force balance between centrifugal $F^x_{\rm cent}$ and adhesion $F_{\rm adh}^x$ forces
then leads to the following expression for the flow deviation:
\begin{equation}
\delta\psi_0 = {\cal F}\left[ {r_i\over {\cal R} }\right]\sqrt{(1+\cos\theta_0)\over {\rm We}}
\label{HC}
\end{equation}
 where ${\rm We}=\rho U^2 e_0/\gamma$ is the Weber number constructed on the film thickness $e_0$, and the geometrical factor ${\cal F}$ takes the simple form ${\cal F} \sim \sqrt{{\cal R}\, r_i/e_0^2}$  for small $e_0$.

 A few comments are in order. 
 First, as announced, this hydro-capillary description does indeed connect the large scale fluid properties to the surface properties: {\it via} its geometry but more interestingly {\it via} its surface properties and contact angle $\theta_0$.
 Furthermore, it fully reproduces all experimental observations in Figs. \ref{fig2}:
 the angle of deviation $\psi_0=\pi/2-\delta\psi_0$ is indeed predicted to increase with the fluid velocity ($U$, or ${\rm We}$), as well as with the contact angle of the surface ($\theta_0$); also $\psi_0$ decreases with the radius of curvature of the spout ($r_i$). 

It is finally interesting to address the dripping and the corresponding threshold velocity, plotted in Fig.\ref{fig3}-b.
While the above argument does not predict intrinsically a limit  of stability for the flow, a criterion in terms
 of a minimal flow deviation $\delta\psi_0^{\rm min}$ in Eq. (\ref{HC}) does 
suggest a corresponding threshold Weber scaling as
\begin{equation}
{\rm We}_c \propto {{\cal R}\, r_i\over e_0^2}(1+\cos\theta_0)
\label{Wec}
\end{equation}
This prediction is compared to the experimental results in Fig. \ref{fig3}, showing again a very good agreement: both the predicted linear scaling on $1+\cos\theta_0$ (Fig. \ref{fig3}-b) and the dependence of its slope on
$r_i$ (Fig. \ref{fig3}-b inset) reproduce the experimental results.
Altogether the hydro-capillary picture is seen to capture the main features of the teapot effect. It
solves accordingly the flow separation question in terms of a novel, capillary meniscus, ingredient.
We finally note that this hydro-capillary picture differs strongly from the -- viscosity-dependent-- splash mechanism in \cite{Duez}:
in contrast to splashes, the capillary meniscus is here stationary and wetting {\it dynamics} is thus not relevant for the dripping mechanism.

\begin{figure}[t]
\begin{center}
\includegraphics[width=8cm,height=!]{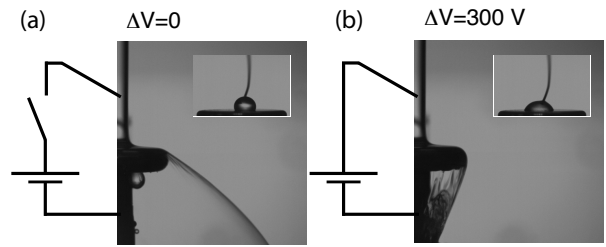}
\caption{{Electro-wetting control of dripping}: A tunable wettability of the surface is achieved using electro-wetting. The contact angle is controled by imposing various electric potential drop $\Delta V$ between the liquid and the solid surface (see inset). A contact angle between
110$^\circ$ and 60$^\circ$ is achieved for $\Delta V$ ranging between 0 and 300V. This leads to 
an active control of the ejection and dripping of the liquid on the impacter: no dripping is obtained for
$\Delta V=0$, while dripping is measured for $\Delta V=300V$. 
\label{fig4}}
\end{center} 
\end{figure}

Beyond this understanding, our results suggest that the flow pattern may be directly controlled {\it via}
a tuning of surface wetting properties. As shown in the recent years \cite{Berge}, electro-wetting is
a very efficient solution to tune  the surface properties: 
the application of an electric potential drop on a polarized surface leads to a direct modification  
of the contact angle \cite{Mugele}. 
We have coupled our dripping geometry in Fig.\ref{fig2} to an electro-wetting
set-up, 
Fig. \ref{fig4}. 
Impactors are covered by a dielectric coating -- a 10$\mu$m thick Parylene C layer --, and then further coated by a thin hydrophobic fluoropolymer AF1600 (Dupont) layer \cite{Welters}. An electric drop $\Delta V$ applied between the liquid and the impacter allows to tune the contact angle on the impacter between $\theta_0=110^\circ$ to $\theta_0=60^\circ$ as $\Delta V$ is varied between 0 and 300 V, see insets of Fig.\ref{fig4} [the voltage dependence of the contact angle was checked to follow a Lippmann equation for small voltages, with a saturation above 200V, not shown]. Now, when a liquid jet impacts the liquid surface,  we observe that the ejection of the fluid 
can be tuned directly -- and dynamically -- by the applied potential drop $\Delta V$. This is illustrated in Fig.\ref{fig4}, where dripping is induced under an applied potential drop (Fig.\ref{fig4}-b), while the fluid is ejected
when this applied potential drop is absent. 
Such an active control opens new application perspectives to dynamically shape flow patterns 
\cite{patent}.

To summarize, we have demonstrated the crucial influence of surface wettability on separation of rapid flows. As a paradigm superhydrophobic surfaces fully avoid dripping, and thus beat the ``teapot effect''. Experimental results are rationalized on the basis of a novel hydro-capillary adhesion phenomenon, coupling inertial flows to a capillary adhesion mechanism. This phenomenon effectively bridges the gap between the small (surface) and large (flow) scales. It opens novel strategies to shape large-scale liquid flows using micro- and nano- engineered surfaces.

This project was supported by DGA. 
We thank Jacques Tardy (INL) for the parylene coatings.

\end{document}